\documentclass[aps,prb,twocolumn,showpacs,groupedaddress]{revtex4-1}
\usepackage[dvips]{graphicx}
\usepackage{amsmath}
\usepackage{bm}
\begin{document}

\title{Frenkel-like plasmonic excitons in plasmonic lattices:\\
Energy spectrum, radiative relaxation, and Bose-Einstein condensation}

\author{V.G.~Bordo}
\email{bordo@mci.sdu.dk}

\affiliation{NanoSyd, Mads Clausen Institute, Syddansk Universitet, Alsion 2, DK-6400 S{\o}nderborg, Denmark}


\date{\today}

\begin{abstract}
The concept of quantum plasmonic excitations in plasmonic lattices, which similarly to Frenkel excitons in molecular crystals propagate by hopping from one nanoparticle to another, is introduced. A consistent quantum theory of such plasmonic excitons, beginning with the quantization of localized surface plasmons in a metal nanoparticle and including the radiative relaxation in both 1D and 2D lattices near a reflective substrate surface, is developed. A possible room-temperature Bose-Einstein condensation in the quantum gas of plasmonic excitons is also discussed.
\end{abstract}


\maketitle

\section{Introduction}
The field of plasmonics has recently received a new, quantum dimension. Quantum plasmonics embraces any quantum effects which emerge in the interaction between light and metal nanostructures. It holds promise for diverse quantum technologies, including quantum computing, quantum cryptography, metrology and sensing \cite{Shalaev11,Jacob12,Park12,Kim13,QP16,Mortensen17,Khurgin17}.\\ 
While the quantum properties of light and atomic light emitters have been thoroughly investigated in quantum optics, the quantum nature of surface plasmons, which are supported by metal sub-wavelength structures, has not yet received a proper attention. In particular, the quantization of surface plasmon modes of spherical metal nanoparticles (NPs) and their arrays - one of the most simple, albeit rigorous, models in plasmonics - has not been comprehensively treated.\\
The analysis of surface plasma oscillations in a metallic sphere was first given by Jensen in 1937 \cite{Jensen37}. He used the hydrodynamical model for an electron gas which oscillates in the field of homogeneously smeared out positive charge of nuclei. Basing on this so-called jellium model, Crowell and Ritchie \cite{Crowell68} first quantized the surface plasmon field to calculate the cross section for surface plasmon creation by fast electrons and the radiative decay rate for dipole surface plasmons. Their paper 
remained, however, some aspects of quantization, which are important for other applications, beyond its scope.\\
The quantum properties of arrays of metal NPs, which are known as plasmonic lattices or plasmonic crystals, have not been explored in detail as well. Linear chains of metal NPs can operate as plasmon waveguides and switching elements, which provide a sub-wavelength energy transport \cite{Quinten98,Krenn99,Atwater00,Maier03,Alu06,Karpov13}. These phenomena are usually described in terms of coupled (collective) plasmon modes which originate from the classical near-field electrodynamic interaction between NPs \cite{Atwater00,Weber04,Polman06}. The coupling strength, which is obtained in such a model and is inherently classical, then is used to construct a Hamiltonian in order to describe the quantum behavior of a metal NP array \cite{Lee12,Jalabert15,Jalabert16}. Although one writes the Hamiltonian in the quantized form, its eigenfunctions are not introduced that obscures the physical sense of the corresponding quantum states. Moreover, the knowledge of the wave functions allows one to calculate the transition matrix elements relevant to different processes which involve a transition between plasmonic states. In particular, they can be used to
obtain the radiative relaxation rates in a very simple manner exploiting Fermi's golden rule.\\ 
The above arguments reveal a need in a consistent quantum approach, which provides both the spectrum of quantum plasmonic states and the corresponding wave functions. In the present paper, we theoretically investigate the energy spectrum of quantum plasmonic excitations in both 1D and 2D spherical metal NPs arrays and their radiative relaxation rates in the vicinity of a reflective substrate surface. We show that such states form excitation waves and their wave packets, which we call "plasmonic excitons", propagate by hopping from one nanoparticle to another, similarly to Frenkel excitons in molecular crystals \cite{Frenkel,Agranovich}. These quasiparticles should not be confused with plasmon-exciton polaritons \cite{Rivas13}, which originate from a coupling between plasmon modes of a lattice, ordinary excitons in the incorporated dye and incident light.\\ 
This view gives a hint that plasmonic excitons, like other Bose  quasiparticles (photons in a cavity, excitons, exciton polaritons, surface plasmon polaritons in plasmonic lattices, etc.), can exhibit collective quantum behavior, in particular Bose-Einstein condensation \cite{Keldysh68,Yamamoto02,West07,Weitz10,Rivas13,Torma18}. In the present paper, we derive analytical expressions for the critical temperature of Bose-Einstein condensation in both 1D and 2D quantum gases of plasmonic excitons. We show that the transition to the condensate phase can occur at room temperature for moderate numbers of quasiparticles.\\
The paper is organized as follows. In Sec. \ref{sec:quantization}, a detailed description of the quantization of multipole plasmon modes of a metal sphere, which provides a basis for the further discussion, is given. In Sec. \ref{sec:excitons}, the plasmonic excitons in both 1D and 2D plasmonic lattices are introduced and their energy spectrum and wave functions are found. In Sec. \ref{sec:relaxation}, the radiative relaxation rates for plasmonic lattices suspended above a reflective surface are calculated. The Bose-Einstein condensation of the quantum gas of plasmonic excitons is discussed in Sec. \ref{sec:BEC}. In Sec. \ref{sec:conclusion}, the main results of the paper are summarized.
\section{Quantization of localized surface plasmons}\label{sec:quantization}
\subsection{Hydrodynamical model}
The motion of electrons is governed by three coupled partial differential equations, written by Crowell and Ritchie \cite{Crowell68} in the form
\begin{eqnarray}
\nabla\ \dot{\psi}({\bf r},t)=-\frac{e}{m}\nabla \phi({\bf r},t)+\frac{\beta^2}{n_0}\nabla n({\bf r},t),\label{eq:force}\\
\nabla^2\phi({\bf r},t)=4\pi en({\bf r},t),\label{eq:Poisson}\\
\nabla^2\psi({\bf r},t)=\frac{1}{n_0}\dot{n}({\bf r},t),\label{eq:continuity}
\end{eqnarray}
which are the force equation, Poisson's equation and continuity equation, respectively. Here $\psi({\bf r},t)$ is the velocity potential which determines the electron velocity through the relation ${\bf v}({\bf r},t)=-\nabla \psi({\bf r},t)$, $\phi ({\bf r},t)$ is the electrostatic potential, $n({\bf r},t)$ is the deviation of the electron density from its equilibrium value $n_0$, $e$ and $m$ are the electron charge and mass, respectively, and a dot above a symbol denotes the partial time derivative. The quantity $\beta$ is the root mean square speed which characterizes the propagation of the electron density disturbance. It can be expressed in terms of the Fermi velocity as $\beta=(3/5)^{1/2}v_F$ \cite{Nakamura83}.\\ 
Equations (\ref{eq:force})-(\ref{eq:continuity}) should be complemented by the boundary conditions which express the continuity of the electrostatic potential along with the normal component of the electric displacement vector at the surface of the sphere, $r=R$, and the vanishing of the normal component of the electron velocity at the surface of the sphere, i.e.
\begin{equation}\label{eq:boundary}
\left(\frac{\partial}{\partial r} \psi(r,t)\right)_{r=R} = 0.\\
\end{equation}
One can notice that the action of the $\nabla$ operator on Eq. (\ref{eq:force}) and taking the time derivative of Eq. (\ref{eq:continuity}) give the same left hand side parts. Equating the corresponding right hand side parts and using Eq. (\ref{eq:Poisson}), one obtains the equation for the electron density disturbance
\begin{equation}
\ddot{n}({\bf r},t)-\beta^2\nabla^2 n({\bf r},t) + \omega_p^2 n({\bf r},t)=0,
\end{equation}
where 
\begin{equation}
\omega_p=\sqrt{\frac{4\pi e^2n_0}{m}}
\end{equation}
is the plasma frequency. Considering the harmonic oscillations in the form $n({\bf r},t)=\rho({\bf r})\cos\omega t$, one comes to the equation for the amplitude $\rho({\bf r})$
\begin{equation}\label{eq:rho}
\nabla^2 \rho({\bf r})-\frac{\omega_p^2-\omega^2}{\beta^2}\rho({\bf r})=0.\\
\end{equation}
The solution of Eq. (\ref{eq:rho}) is found by means of expansion in terms of the spherical harmonics, $Y_{lm}(\theta,\varphi)$, as follows
\begin{equation}\label{eq:expansion}
\rho(r,\theta,\varphi)=\sum_{lm}A_{lm}j_l[i(\omega_p^2-\omega^2)^{1/2}r/\beta ]Y_{lm}(\theta,\varphi),
\end{equation}
where $j_l(x)$ is the spherical Bessel function of the first kind of order $l$ and $A_{lm}$ are the coefficients which can be found from the initial conditions. This expansion decomposes the collective electron oscillations into different modes specified by the integers $l$ and $m$. We assume here and in what follows that $Y_{lm}(\theta,\varphi)$ are the real, normalized spherical harmonics; the values $0\leq m\leq l$ correspond to the even functions of $\varphi$, while the values $-l\leq m < 0$ correspond to the odd functions of $\varphi$ \cite{Morse53}. The implementation of the boundary conditions leads to the dispersion relations given in Ref. \cite{Crowell68} which determine the allowed frequencies $\omega_l$ for each mode.\\
The argument of the spherical Bessel function in Eq. (\ref{eq:expansion}) dictates the characteristic length scale $\delta_l=\beta/(\omega_p^2-\omega_l^2)^{1/2}$. If $\delta_l\ll R$ one can use the asymptotic expansion of the functions $j_l(x)$ in the dispersion relation. Assuming that $\omega_l\sim \omega_p$ and taking for an estimate $\omega_p\sim 10^{16}$ s$^{-1}$ and $v_F\sim 10^8$ cm/s, one obtains $\delta_l\sim 1 \text{\AA}$. For large enough spheres such that $R\gg 1 \text{\AA}$ the dispersion relation is reduced to $\omega_l^2=\omega_p^2l/[\epsilon_m(l+1)+l]$ with $\epsilon_m$ being the dielectric constant of the medium surrounding the sphere \cite{Crowell68}. In particular, for the dipole plasmon mode ($l=1$) $\omega_1=\omega_p/(2\epsilon_m+1)^{1/2}$.\\
Under the same conditions the amplitude $\rho({\bf r})$ for points not very close to the center of the sphere takes the form
\begin{equation}
\rho(r,\theta,\varphi)\approx \frac{1}{2r}\sum_{lm}i^l\delta_le^{R/\delta_l}e^{-(R-r)/\delta_l}A_{lm}Y_{lm}(\theta,\varphi).
\end{equation}
This quantity drops very rapidly (on the scale of 1 $\text{\AA}$) with the distance from the surface of the sphere that allows one to approximate the radial dependence of the electron density disturbance by Dirac's delta function as follows
\begin{equation}\label{eq:surface}
n(r,\theta,\varphi)\approx \delta (r-R)\sigma (\theta,\varphi),
\end{equation}
where
\begin{equation}\label{eq:sigma}
\sigma (\theta,\varphi)=\sum_{lm}\sigma_{lm}Y_{lm}(\theta,\varphi)
\end{equation}
is the surface density of the electron disturbance and $\sigma_{lm}$ are constants related with $A_{lm}$.\\
In this approximation, the electrostatic potential both inside and outside the sphere obeys the Laplace equation $\nabla^2 \phi({\bf r}) =0$, whereas the jump of the normal component of the electric displacement vector at the surface of the sphere is equal to $4\pi e\sigma(\theta,\varphi)$. The solution of this standard electrostatic problem gives
\begin{equation}\label{eq:in}
\phi_{in}({\bf r})=\sum_{lm}\left(\frac{r}{R}\right)^l \phi_{lm}Y_{lm}(\theta,\varphi)
\end{equation}
for the potential inside the sphere and
\begin{equation}\label{eq:out}
\phi_{out}({\bf r})=\sum_{lm}\left(\frac{R}{r}\right)^{l+1}\phi_{lm} Y_{lm}(\theta,\varphi)
\end{equation}
for the potential outside it with
\begin{equation}
\phi_{lm}=-\frac{4\pi eR}{\epsilon_m(l+1)+l}\sigma_{lm}.\\
\end{equation}
Let us integrate now the equation $\dot{\psi}=-(e/m)\phi+(\beta^2/n_0)n$, which follows from Eq. (\ref{eq:force}), over a small radial interval $(R-\delta,R+0)$, where $\delta\sim \beta/\omega_p$ is the length scale within which the electron density disturbance is localized. Taking into account Eq. (\ref{eq:surface}), one obtains
\begin{equation}\label{eq:delta}
\delta \dot{\psi}_R(\theta,\varphi) \approx  -\delta\frac{e}{m}\phi_R(\theta,\varphi) + \frac{\beta^2}{n_0}\sigma(\theta,\varphi),
\end{equation}
where
\begin{equation}\label{eq:psiR}
\psi_R(\theta,\varphi)=\sum_{lm}\psi_{lm}Y_{lm}(\theta,\varphi)
\end{equation}
and
\begin{equation}
\phi_R(\theta,\varphi)=\sum_{lm}\phi_{lm}Y_{lm}(\theta,\varphi)
\end{equation}
are the values of $\psi({\bf r})$ and $\phi({\bf r})$, respectively, at $r=R$. Equation (\ref{eq:delta}) is reduced to the following partial equations
\begin{equation}
\delta\dot{\psi}_{lm}\approx \frac{\beta^2}{n_0}\left\{\frac{R}{\delta[\epsilon_m(l+1)+l]}+1\right\}\sigma_{lm}.
\end{equation}
For not very large $l$, such that $\epsilon_m(l+1)+l\ll R/\delta$, the unity in the curly brackets can be neglected that leads to the approximate relation
\begin{equation}\label{eq:psidot}
\dot{\psi}_{lm}\approx \frac{\omega_l^2R}{ln_0}\sigma_{lm}.
\end{equation}
Assuming that $\sigma_{lm}\sim \cos\omega_lt$ and $\psi_{lm}\sim \sin\omega_lt$ one finds from here
\begin{equation}\label{eq:psi}
\psi_{lm}\approx -\frac{R}{ln_0}\dot{\sigma}_{lm}.\\
\end{equation}
\subsection{Hamiltonian}
The Hamiltonian of electrons has the form \cite{Crowell68}
\begin{equation}
H=-\frac{1}{2} \int \left(mn_0\psi\nabla^2\psi + e\phi n - \frac{m\beta^2}{n_0}n^2\right) d{\bf r},
\end{equation}
where the term corresponding to the kinetic energy has been obtained using Green's theorem and Eq. (\ref{eq:boundary}). This term can be further transformed with the use of Eq. (\ref{eq:continuity}), while the sum of the other two terms can be simplified with the use of Eq. (\ref{eq:force}). As a result one obtains
\begin{equation}
H=\frac{m}{2}\int (n\dot{\psi}-\psi\dot{n})d{\bf r}=\frac{m}{2}R^2\int(\sigma\dot{\psi}_R-\psi_R\dot{\sigma})d\Omega,
\end{equation}
where $d\Omega = \sin\theta d\theta d\varphi$ and we have used Eq. (\ref{eq:surface}). Substituting here expansions (\ref{eq:sigma}) and (\ref{eq:psiR}), one comes to the expression
\begin{equation}
H=\frac{m}{2}R^2\sum_{lm}(\sigma_{lm}\dot{\psi}_{lm}-\psi_{lm}\dot{\sigma}_{lm}),
\end{equation}
where we have used the orthonormality of the spherical harmonics. Finally, with the use of Eqs. (\ref{eq:psidot}) and (\ref{eq:psi}) the Hamiltonian takes the form
\begin{equation}\label{eq:oscillator}
H=\frac{1}{2}\sum_{lm}(P_{lm}^2+\omega_l^2 Q_{lm}^2),
\end{equation}
where
\begin{equation}\label{eq:Q}
Q_{lm}=\sqrt{\frac{mR^3}{ln_0}}\sigma_{lm}
\end{equation}
and $P_{lm}=\dot{Q}_{lm}$ are the canonical plasmonic field variables - the generalized coordinates and momenta, respectively, associated with the plasmonic mode $\{lm\}$ \cite{Landau1}.\\
Equation (\ref{eq:oscillator}) provides the expansion of the Hamiltonian of electrons in terms of the Hamiltonians of elementary harmonic oscillators of the plasmonic field. To make the transition to its quantum description one has to consider the canonical variables as operators satisfying the commutation rules \cite{Landau4}
\begin{equation}
[\hat{P}_{lm},\hat{Q}_{l^{\prime}m^{\prime}}]=-i\hbar\delta_{ll^{\prime}}\delta_{mm^{\prime}}.\\
\end{equation}
The annihilation and creation operators of plasmonic quanta in the mode $\{lm\}$ are introduced as 
\begin{equation}\label{eq:a}
\hat{a}_{lm}=\frac{1}{\sqrt{2\hbar\omega_l}}\left(\omega_l\hat{Q}_{lm}+i\hat{P}_{lm}\right)
\end{equation}
and
\begin{equation}\label{eq:a+}
\hat{a}^{\dagger}_{lm}=\frac{1}{\sqrt{2\hbar\omega_l}}\left(\omega_l\hat{Q}_{lm}-i\hat{P}_{lm}\right),
\end{equation}
respectively. Then the normally ordered quantized Hamiltonian takes the form
\begin{equation}\label{eq:hamiltonian}
\hat{H}=\sum_{lm}\hbar\omega_l\hat{a}^{\dagger}_{lm}\hat{a}_{lm}.
\end{equation}
Its eigenvalues are given by
\begin{equation}
E=\sum_{lm}N_{lm}\hbar\omega_l,
\end{equation}
where $N_{lm}$ is the number of plasmonic quanta in the mode $\{lm\}$. \\
Taking into account Eqs. (\ref{eq:a}), (\ref{eq:a+}) and (\ref{eq:Q}), one finds 
\begin{equation}
\hat{\sigma}_{lm}=\sqrt{\frac{l\hbar n_0}{2m\omega_l R^3}}\left(\hat{a}_{lm}+\hat{a}^{\dagger}_{lm}\right).
\end{equation}
The operators of the electrostatic potential inside and outside the sphere, $\hat{\phi}_{in}$ and $\hat{\phi}_{out}$, are obtained from Eqs. (\ref{eq:in}) and (\ref{eq:out}), respectively, by means of the substitution
\begin{equation}
\hat{\phi}_{lm}=-\sqrt{\frac{2\pi l\hbar \omega_l}{[\epsilon_m(l+1)+l]R}}\left(\hat{a}_{lm}+\hat{a}^{\dagger}_{lm}\right).
\end{equation}
In particular, the contribution of the dipole plasmonic mode ($l=1$) of the sphere suspended in vacuum ($\epsilon_m=1$) to the electrostatic potential is identical with the result given by Crowell and Ritchie \cite{Crowell68}.\\
It is also of interest to calculate the operator of the dipole moment, $\hat{\bf p}$, of the sphere which we will use later on. Taking the direction of the dipole moment specified by the unit vector $\hat{\bf e}$ as the $z$ axis, one finds
\begin{eqnarray}\label{eq:dipole}
\hat{\bf p}=\hat{\bf e}e\int z\hat{\rho}({\bf r})d{\bf r}=\hat{\bf e}\sqrt{\frac{4\pi}{3}}eR^3\hat{\sigma}_{10}\nonumber\\
=\hat{\bf e}\sqrt{\frac{2\epsilon_m+1}{6}}\sqrt{\hbar \omega_1R^3}\left(\hat{a}_{10}+\hat{a}^{\dagger}_{10}\right).
\end{eqnarray}
\subsection{Interaction of two spheres}\label{sec:two}
The operator of the electrostatic interaction between two identical spheres can be found as
\begin{eqnarray}
\hat{V}=-\frac{e}{2}\int\left(\hat{\rho}^{(1)}\hat{\phi}_{out}^{(2)}+\hat{\rho}^{(2)}\hat{\phi}_{out}^{(1)}\right)d{\bf r}\nonumber\\
=-\frac{e}{2}R^2\int\left(\hat{\sigma}^{(1)}\hat{\phi}_{out}^{(2)}+\hat{\sigma}^{(2)}\hat{\phi}_{out}^{(1)}\right)d\Omega,
\end{eqnarray}
where the superscripts $(1)$ and $(2)$ refer to the first and second sphere, respectively. Assuming that the radius of the spheres is much smaller than the distance between them, $d$, and expanding the interaction potential in terms of the ratio $R/d$ keeping the lowest non-vanishing terms, one obtains
\begin{eqnarray}
\hat{V}=\frac{1}{2}\left(\frac{R}{d}\right)^3\hbar \omega_1(1-3\cos^2\theta_0)\nonumber\\
\times\left(\hat{a}^{(1)}_{10}+\hat{a}^{(1)\dagger }_{10}\right)\left(\hat{a}^{(2)}_{10}+\hat{a}^{(2)\dagger }_{10}\right),
\end{eqnarray}
where we have assumed that only the dipole mode $\{10\}$ can be populated in both spheres and $\theta_0$ is the angle between the quantization axis and the line connecting the centers of the spheres. This operator can be rewritten in terms of the dipole moment operator, Eq. (\ref{eq:dipole}), as follows
\begin{equation}
\hat{V}=\frac{3}{2\epsilon_m+1}\frac{1}{d^3}\left[\hat{\bf p}^{(1)}\cdot\hat{\bf p}^{(2)}-3(\hat{\bf p}^{(1)}\cdot{\bf n})(\hat{\bf p}^{(2)}\cdot{\bf n})\right],
\end{equation}
where ${\bf n}$ is the unit vector directed along the line between the centers of two spheres.\\
The Hamiltonian of two interacting spheres has the form
\begin{equation}
\hat{H}=\hat{H}^{(1)}+\hat{H}^{(2)}+\hat{V},
\end{equation}
where $\hat{H}^{(1)}$ and $\hat{H}^{(2)}$ are the Hamiltonians of isolated spheres given by Eq. (\ref{eq:hamiltonian}). The unperturbed Hamiltonian, $\hat{H}^{(1)}+\hat{H}^{(2)}$, has a twofold degenerate eigenvalue $E_0=\hbar\omega_1$ which corresponds to a single dipole plasmon quantum in the system located either at one sphere (the state $\psi_1=\mid 10\rangle$) or another (the state $\psi_2=\mid 01\rangle$). The correct wave functions in the zeroth approximation are linear combinations of the form \cite{Landau3}
\begin{equation}
\Psi=c_1\psi_1 + c_2\psi_2,
\end{equation}
where the coefficients $c_1$ and $c_2$ are found from the secular equation and we assume that the wave functions $\psi_1$ and $\psi_2$ are normalized to unity. The first-order corrections to the eigenvalue are given by $E_1=\pm \mid V_{12}\mid$, where
\begin{equation}\label{eq:V12}
V_{12}=V_{21} =\frac{1}{2}\left(\frac{R}{d}\right)^3\hbar \omega_1 (1-3\cos^2\theta_0) ,
\end{equation}
i.e. the energy spectrum of the dipole plasmon modes displays a splitting of magnitude $2\mid V_{12}\mid$. \\
The solution of the secular equation reveals \cite{Landau3} that if at the initial instant the plasmon quantum is localized at one sphere, the probability to find it at another sphere will vary periodically with time, with frequency $\omega_0=2\mid V_{12}\mid/\hbar$. This consideration does not take, however, into account the relaxation processes. Nevertheless it provides an adequate description if the relaxation time, $\tau$, is longer than the period of such oscillations $T=2\pi/\omega_0$. Alternatively, the relaxation rate, $\tau^{-1}$, should be less that the frequency of oscillations. Taking for an estimate $\tau\sim 10^{-12}$ s \cite{Nisoli97} and $2\pi/\omega_1\sim 10^{-15}$ s, one finds that this criterion is fulfilled if $R/d>0.1$.\\
\subsection{Radiative relaxation in a sphere}
The formalism developed above allows one to apply the results obtained for radiative processes in a two-level quantum system. Let us consider the rate of the radiative relaxation in a metallic sphere in which the dipole plasmonic mode is populated with a single quantum. Such a process can be regarded as spontaneous annihilation of a plasmon and simultaneous creation of a photon due to the interaction with the electromagnetic vacuum of the surrounding medium. In the dipole approximation, which is valid if the sphere diameter is much smaller than the wavelength of interest, its rate is found from Fermi's golden rule as \cite{Landau4}
\begin{equation}\label{eq:w}
w=\frac{4\sqrt{\epsilon_m}\omega_1^3}{3\hbar c^3}\mid {\bf p}^{fi}\mid ^2,
\end{equation}
where $c$ is the speed of light in vacuum, ${\bf p}^{fi}$ is the transition dipole moment between the initial and final states and we have taken into account both the reduction of the speed of light and the renormalization of the electric field creation and annihilation operators in a dielectric \cite{Bhat88}.\\
In our case the initial state is the state with a single dipole plasmon quantum, $\mid 1\rangle$, while the final state is the vacuum state of the plasmonic field, $\mid 0\rangle$. The substitution of the dipole moment operator, Eq. (\ref{eq:dipole}), into Eq. (\ref{eq:w}) gives
\begin{equation}\label{eq:w1}
w=\frac{2}{9}\sqrt{\epsilon_m}(2\epsilon_m+1)\frac{\omega_1^4R^3}{c^3}.
\end{equation}
This expression is identical with the result obtained in Ref. \cite{Melikyan04} from a classical consideration and it coincides with the formula given in Ref. \cite{Crowell68} for $\epsilon_m=1$. For example, for an Ag sphere of radius $R=20$ nm embedded into a medium with $\epsilon_m=2.25$ ($\omega_1=6.0\times 10^{15}$ s$^{-1}$) one obtains $w\approx 7.0\times 10^{14}$ s$^{-1}$.\\
Let us note that if in the initial state the dipole plasmon mode is populated with $\mathcal{N}$ quanta, the radiative relaxation (plasmon annihilation) rate is given by $\mathcal{N}w$, i.e. it is a process stimulated by plasmons. \\
\section{Plasmonic excitons}\label{sec:excitons}
The results highlighted in the previous section allow one to develop a quantum description of plasmonic excitations in plasmonic lattices. We assume that the distance between NPs is large enough so that no electron tunneling is possible between them. We are interested in the lowest excited state of such a system in which one NP is excited (i.e. its dipole plasmon mode is populated with a single quantum) and the others are non-excited. This objective resembles the problem of finding excited states of a molecular crystal discussed first by Frenkel \cite{Frenkel} and can be treated in a similar way. In what follows, we consider 1D lattices (linear chains) and rectangular 2D lattices of NPs.\\
\subsection{1D lattice}\label{sec:1D}
One can generalize the results obtained in Sec. \ref{sec:two} for two NPs to a linear chain of $N$ identical NPs separated by the distance $a$ from each other. This time the excited state is $N$-fold degenerate and different eigenstates $\psi_i$ of the unperturbed Hamiltonian correspond to the excitation localized at the $i$-th NP. The correct wave functions in the zeroth approximation have the form
\begin{equation}
\Psi=\sum_{i=1}^Nc_i\psi_i.
\end{equation}
Assuming that only the interaction between neighboring NPs is essential\cite{Note2}, one obtains the matrix of the interaction operator in the basis of functions $\psi_i$ in an $N\times N$ tridiagonal Toeplitz form
\begin{equation}\label{eq:toeplitz}
\hat{V}=
\begin{pmatrix} 
0 & V_{12} & 0 & \cdots & 0 & 0 \\ 
V_{21} & 0 & V_{12} & \cdots & 0 & 0 \\ 
0 & V_{21} & 0 & \cdots & 0 & 0 \\ 
\vdots & \vdots & \vdots & \ddots & \vdots & \vdots \\
0 & 0 & 0 & \cdots & 0 & V_{12} \\ 
0 & 0 & 0 & \cdots & V_{21} & 0 
\end{pmatrix}.
\end{equation}
The eigenvalues of this matrix, which determine the first-order corrections to the unperturbed energy $E_0=\hbar\omega_1$, are found as\cite{Reichel13}
\begin{equation}\label{eq:E1n}
E_{1n}=2 V_{12} \cos\left(\frac{n\pi}{N+1}\right),
\end{equation}
where $V_{12}$ is given by Eq. (\ref{eq:V12}) with $d$ replaced by $a$ and $n=1,2,...,N$. For a large number of NPs, $N$, the eigenvalues form a quasi-continuous excitonic band 
\begin{equation}\label{eq:Ek}
E(k)=E_0+2V_{12}\cos ka
\end{equation}
disposed between the energies $E_0-2\mid V_{12}\mid $ and $E_0+2\mid~V_{12}\mid $.\\
\begin{figure}
\includegraphics[width=\linewidth]{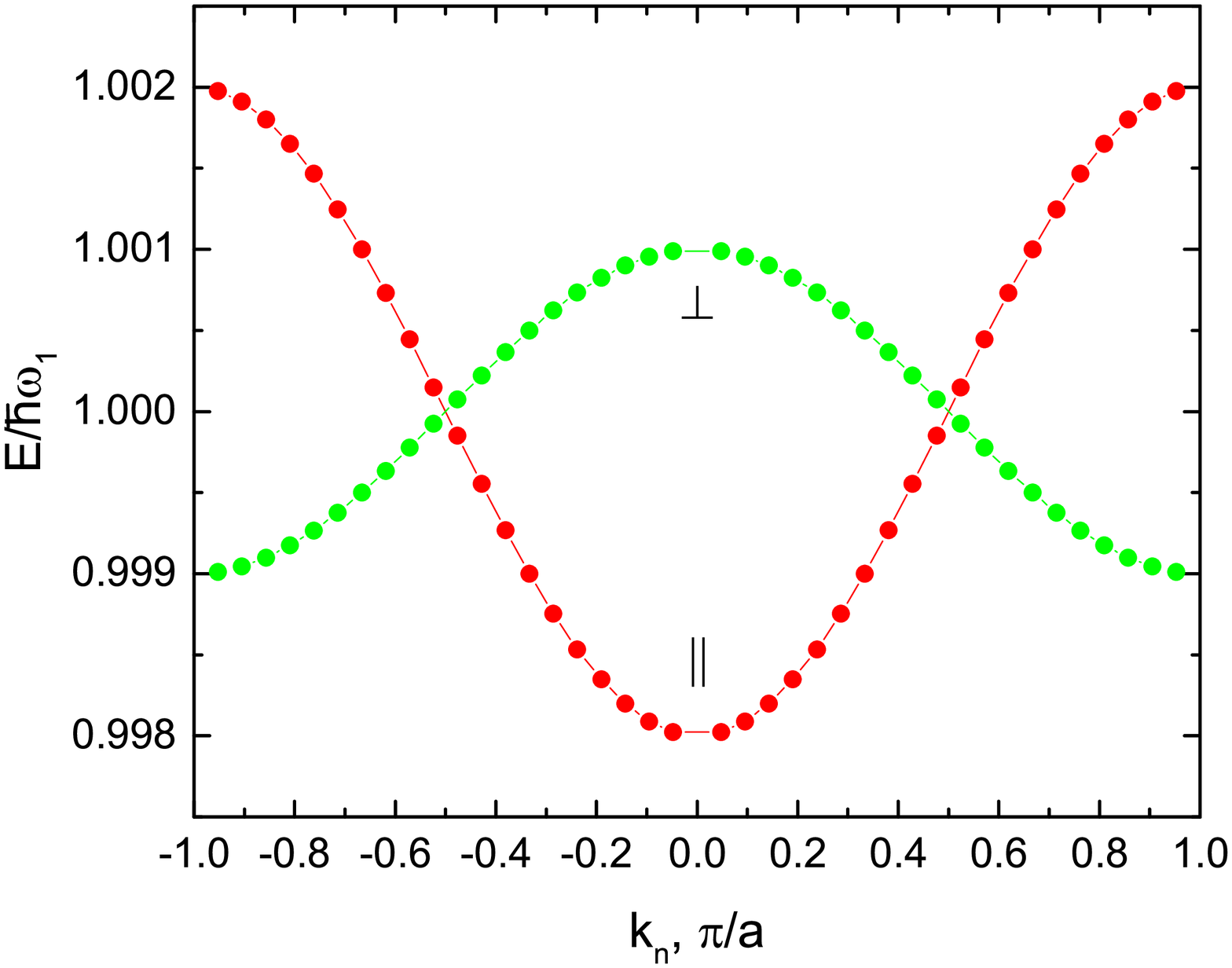}
\caption{\label{fig:energy} The plasmonic exciton energy spectrum (shown by dots) normalized to $E_0=\hbar\omega_1$ for the NP dipole oscillations parallel ($\parallel $) and perpendicular ($\perp $) to the linear chain of NPs. (The negative values of $k_n$ correspond to the opposite direction of propagation.) $R/a=0.1$ and $N=20$.}
\end{figure}
The components of the eigenvectors $\vec{c}_n=(c_{n1},c_{n2},...,c_{nN})^T$ associated with the eigenvalues $E_{1n}$ have the form \cite{Reichel13}
\begin{equation}
c_{nj}=\sin\left(\frac{nj\pi}{N+1}\right),\quad n,j=1,2,...,N.
\end{equation}
The corresponding wave functions
\begin{equation}
\Psi_n=A_n\sum_{j=1}^Nc_{nj}\psi_j e^{(-i/\hbar)E_nt}
\end{equation}
with $A_n$ being the normalization factor and $E_n=E_0+E_{1n}$ can be regarded as standing waves of the probability amplitude which are composed of two counter-propagating waves
\begin{equation}\label{eq:waves}
\Psi_{n\pm}=A_{n\pm}\sum_{j=1}^N \psi_j e^{\pm ik_nx_j}e^{(-i/\hbar)E_nt},
\end{equation}
where the wave vector of the waves is defined as 
\begin{equation}
k_n=\frac{n\pi}{(N+1)a}
\end{equation}
and the coordinate $x_j=ja$ specifies the position of the $j$-th NP in the chain.\\ 
The solutions (\ref{eq:waves}) have a sense of the excitation waves with the energy $E_n$ and the wave vectors $\pm k_n$. If the plasmonic excitation is initially localized at a certain NP, the corresponding wave function is represented by a wave packet of such excitation waves which propagates with the group velocity 
\begin{equation}
v=\frac{1}{\hbar}\frac{\partial E_n}{\partial k_n}=-\frac{2V_{12}a}{\hbar}\sin k_na.
\end{equation}
It can be either positive or negative, depending on the sign of $V_{12}$ which is dictated by the polarization of the dipole plasmon oscillations [see Eq. (\ref{eq:V12})]. In other words, the direction of the plasmonic exciton propagation can be controlled by the polarization of the external electromagnetic field which excites it.\\
\subsection{2D lattice}
Let us consider now a 2D rectangular array of identical NPs, which contains $N_1$ NPs along the $x$ axis and $N_2$ NPs along the $y$ axis, with an elementary cell of size $a\times b$. Such a structure can be considered as interacting linear chains of NPs, which have been discussed in Sec. \ref{sec:1D}.\\
The Hamiltonian of the system can be written in the form
\begin{equation}
\hat{H}=\sum_{i=1}^{N_2}\hat{\mathcal{H}}_i+\sum_{\substack{i=1\\j=i\pm 1}}^{N_2}\hat{\mathcal{V}}_{ij},
\end{equation}
where $\hat{\mathcal{H}}_i$ is the Hamiltonian of the $i$-th linear chain and $\hat{\mathcal{V}}_{ij}$ is the operator of the interaction between the $i$-th and $j$-th chains in assumption of interaction between neighboring NPs. The unperturbed Hamiltonian, $\sum_{i=1}^{N_2}\hat{\mathcal{H}}_i$, has $N_2$-fold degenerate eigenvalues $E_0+E_{1n}$ [see Eq. (\ref{eq:E1n}) with $N$ replaced by $N_1$]. Taking the correct wave functions in the zeroth approximation in the form
\begin{equation}
\Phi_n=\sum_{i=1}^{N_2}d_{in}\Psi_{in},
\end{equation}
where $\Psi_{in}$ are the normalized eigenfunctions of the Hamiltonian $\hat{\mathcal{H}}_i$, one obtains an $N_2\times N_2$ matrix of the operator $\hat{\mathcal{V}}$ in a tridiagonal Toeplitz form as before [see Eq. (\ref{eq:toeplitz})]. Its eigenvalues
\begin{equation}\
E^{\prime}_{1n^{\prime}}=2 V^{\prime}_{12} \cos\left(\frac{n^{\prime}\pi}{N_2+1}\right),
\end{equation}
where $V^{\prime}_{12}$ is given by Eq. (\ref{eq:V12}) with $d$ replaced by $b$ and $n^{\prime}=1,2,...,N_2$, provide the first-order corrections to the energies of non-interacting chains of NPs.\\
Finally, the plasmonic exciton spectrum is given by
\begin{eqnarray}
E_{nn^{\prime}}=E_0+2 V_{12} \cos\left(\frac{n\pi}{N_1+1}\right)\nonumber\\
+2 V^{\prime}_{12} \cos\left(\frac{n^{\prime}\pi}{N_2+1}\right),
\end{eqnarray}
where $n=1,2,...,N_1$ and $n^{\prime}=1,2,...,N_2$. The corresponding eigenfunctions have the form
\begin{eqnarray}
\Phi_{nn^{\prime}}=A_{nn¨{\prime}}\sum_{j=1}^{N_1}\sum_{k=1}^{N_2}\sin\left(\frac{nj\pi}{N_1+1}\right)\sin\left(\frac{n^{\prime}k\pi}{N_2+1}\right)\nonumber\\
\times\psi_{jk} e^{(-i/\hbar)E_{nn^{\prime}}t},
\end{eqnarray}
where the wave function $\psi_{jk}$ describes an excitation localized at the $j$-th NP in the $k$-th chain. They can be represented as a superposition of propagating probability waves with the wave vectors ${\bf k}_{nn^{\prime}}=k_n\hat{\bf e}_x+k_{n^{\prime}}\hat{\bf e}_y$ with $k_n=n\pi/(N_1+1)a$ and $k_{n^{\prime}}=n^{\prime}\pi/(N_2+1)b$, and $\hat{\bf e}_x$ and $\hat{\bf e}_y$ being the unit vectors along the $x$ and $y$ axes, respectively.\\
In the case of a 2D plasmonic lattice, the group velocity of the plasmonic excitons depends on the direction of their propagation. Its components along the two lattice axes are given by
\begin{equation}
v_x=-\frac{2V_{12}a}{\hbar}\sin k_na
\end{equation}
and
\begin{equation}
v_y=-\frac{2V^{\prime}_{12}b}{\hbar}\sin k_{n^{\prime}}b.\\
\end{equation}
\section{Radiative relaxation}\label{sec:relaxation}
The knowledge of the plasmonic exciton wave function allows one to calculate its radiative relaxation rate through Fermi's golden rule. This problem can be put in a more general context of NP arrays radiating in the vicinity of a reflective surface. In this section, we shall consider both 1D and 2D plasmonic lattices arranged parallel to the substrate surface.\\
In the dipole approximation, the interaction of a system of $N$ NPs with the radiation field is described by the operator
\begin{equation}\label{eq:HI}
\hat{H}_I=-\sum_{j=1}^N \hat{\bf p}_j\cdot \hat{\bf D}({\bf r}_j),
\end{equation}
where $\hat{\bf p}_j$ is the dipole moment operator of the $j$-th NP located at ${\bf r}_j$ and $\hat{\bf D}({\bf r})$ is the transverse displacement field operator. The transition rate from an initial state $\mid i\rangle$ of the system to its final state $\mid f\rangle$ due to the interaction (\ref{eq:HI}) is found as  \cite{Sipe84}
\begin{equation}\label{eq:Rfi}
R_{fi}=\frac{2}{\hbar}\sum_{j,k=1}^N\sum_{\alpha,\beta}p^{fi}_{j\alpha}p^{if}_{k\beta}\text{Im}G_{\alpha\beta}({\bf r}_j,{\bf r}_k;\omega_{fi}),
\end{equation}
where $G_{\alpha\beta}({\bf r},{\bf r}^{\prime};\omega)$ is the Fourier transform of the field correlation function and $\omega_{fi}$ is the transition frequency. The quantity $G_{\alpha\beta}({\bf r},{\bf r}^{\prime};\omega)$ can be identified with the expectation value of the displacement field at ${\bf r}$ generated by a classical dipole, oscillating at frequency $\omega$, located at ${\bf r}^{\prime}$. It can be split into two parts,
\begin{equation}
G_{\alpha\beta}=G_{\alpha\beta}^0+G_{\alpha\beta}^R,
\end{equation}
where the first term originates from the direct dipole field and the second term is due to the dipole field reflected from the surface. Both quantities can be expressed in terms of their spatial Fourier transforms over the lateral coordinates along the surface,
\begin{equation}
G_{\alpha\beta}^{0,R}({\bf r},{\bf r}^{\prime};\omega)=\frac{1}{(2\pi )^2}\int G_{\alpha\beta}^{0,R}(z,z^{\prime};{\bm\kappa},\omega)e^{i{\bm\kappa}\cdot ({\bm\rho}-{\bm\rho}^{\prime})}d{\bm \kappa},
\end{equation}
where ${\bf r}=({\bm \rho},z)$, the $z$ axis is directed along the normal to the surface and the explicit forms of the tensors $\bar{G}^{0,R}(z,z^{\prime};{\bm\kappa},\omega)$ are given in Appendix \ref{sec:apA}.\\
\subsection{1D lattice}
\begin{figure}
\includegraphics[width=\linewidth]{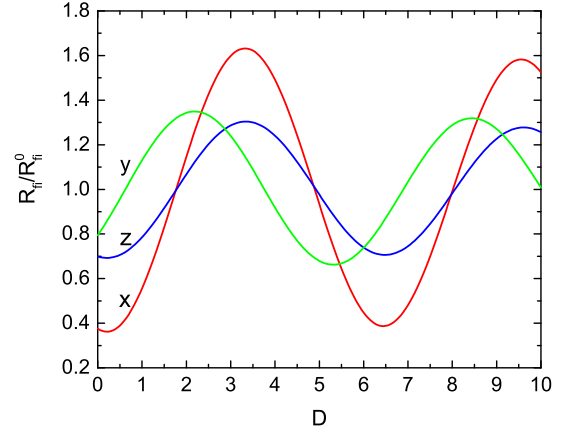}
\caption{\label{fig:rate_d} The radiative relaxation rate in a 1D lattice of Ag NPs at an Ag substrate as a function of the dimensionless distance $D=\tilde{\omega}_{fi}z_0$. The rate is normalized to its value at infinite distance from the substrate for the same polarization. The letters $x$, $y$, and $z$ indicate the direction of the dipole oscillations in NPs. The calculations carried out for the dimensionless wave vector $K=k/\tilde{\omega}_{fi}=0.85$ and $\epsilon_m=2.25$.}
\end{figure}

\begin{figure}
\includegraphics[width=\linewidth]{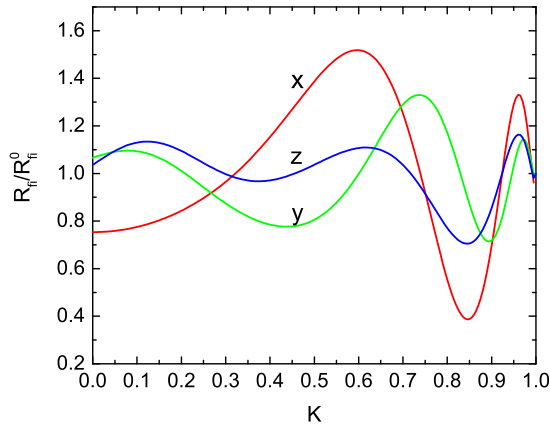}
\caption{\label{fig:rate_k} Same as in Fig. \ref{fig:rate_d}, but as a function of the dimensionless wave vector $K$ calculated for $D=6.5$.}
\end{figure}

We consider the relaxation rate of the propagating probability waves of the form (\ref{eq:waves})
\begin{equation}
\Psi_n=\frac{1}{\sqrt{N}}\sum_{j=1}^N \psi_j e^{ik_nx_j}e^{(-i/\hbar)E_nt},
\end{equation}
where the prefactor ensures the normalization of the wave function to unity. In the course of the radiative transition this wave function plays a role of the initial state, while the final state is the vacuum plasmonic state in which all NPs are non-excited. Calculating the transition dipole moments and substituting them in Eq. (\ref{eq:Rfi}), one finds (see Appendix \ref{sec:apB} for detail)
\begin{eqnarray}\label{eq:Ralpha}
R_{fi}^{\alpha}\approx\frac{(2\epsilon_m+1) \omega_1R^3}{3a}\theta\left(\frac{\tilde{\omega}_{fi}}{k_n}\right)\nonumber\\
\times\int \text{Im}G_{\alpha\alpha}(z_0,z_0;{\bm \kappa}_n,\omega_{fi})d\kappa_y,
\end{eqnarray}
where $\alpha=x,y,z$ specifies the orientation of the dipole moments in NPs, $\tilde{\omega}_{fi}=(\omega_{fi}/c)\sqrt{\epsilon_m}$, $z_0$ is the distance between the chain of NPs and the surface and  ${\bm \kappa}_n=k_n\hat{\bf e}_x+\kappa_y\hat{\bf e}_y$. Here the unit step function $\theta(x)$ indicates that the radiative relaxation is inhibited if the wave vector of the plasmonic exciton is in the non-radiative region $k_n>(\omega_{fi}/c)\sqrt{\epsilon_m}$.\\
Equation (\ref{eq:Ralpha}) determines in particular the radiative relaxation rate in a lattice suspended far from the substrate if one substitutes $G_{\alpha\alpha}^0$ instead of $G_{\alpha\alpha}$.
For the dipoles excited either along the chain of NPs, or perpendicular to it one obtains
\begin{equation}
R_{fi}^0=\frac{3\pi c^3w}{4\sqrt{\epsilon_m}a\omega_1^3}\left(\tilde{\omega}_{fi}^2\pm k_n^2\right)\theta\left(\frac{\tilde{\omega}_{fi}}{k_n}\right),
\end{equation}
where the upper and lower signs correspond to the dipoles oriented perpendicularly and parallel to the chain, respectively, $w$ is the radiative relaxation rate for a single NP, Eq. (\ref{eq:w1}). This result for the perpendicular orientation of dipoles agrees with the radiative damping rate obtained in Ref. \cite{Jalabert16} for the collective plasmon modes in the limit of an infinite chain of NPs taking into account that $\omega_{fi}\approx \omega_1$ and $\epsilon_m=1$. It differs, however, by a factor of two from those calculations for the parallel orientation of dipoles. \\
Figures \ref{fig:rate_d} and \ref{fig:rate_k} show the radiative relaxation rate in a 1D lattice normalized to its value far from the substrate surface as a function of the lattice-surface distance and the wave vector of the exciton, respectively. Both dependencies display oscillating behavior which originates from the interference between the field radiated by the lattice and the one reflected from the surface.\\
\subsection{2D lattice}
In a 2D lattice, as an initial plasmonic state $\mid i\rangle$ we consider the propagating probability waves
\begin{equation}
\Phi_{nn^{\prime}}=\frac{1}{\sqrt{N_1N_2}}\sum_{j=1}^{N_1}\sum_{k=1}^{N_2}\psi_{jk}e^{i{\bf k}_{nn^{\prime}}\cdot {\bm \rho}_{jk}}e^{(-i/\hbar)E_{nn^{\prime}}t},
\end{equation}
where the position of a NP in the 2D array is specified by the vector ${\bm \rho}_{jk}=ja\hat{\bf e}_x+kb\hat{\bf e}_y$. In this case the transition rate for the dipole moments oriented along $\hat{\bf e}_{\alpha}$ is given by
\begin{equation}
R_{fi}^{\alpha}\approx\frac{(2\epsilon_m+1)\omega_1R^3}{3\pi ab}\theta\left(\frac{\tilde{\omega}_{fi}}{k_{nn^{\prime}}}\right)\text{Im}G_{\alpha\alpha}(z_0,z_0;{\bf k}_{nn^{\prime}},\omega_{fi}).
\end{equation}
In particular, for a 2D lattice suspended in medium with the dielectric function $\epsilon_m$ one finds
\begin{equation}
R_{fi}^{\alpha 0}=\frac{3c^3w\tilde{\omega}_{fi}^2}{\sqrt{\epsilon_m}ab\omega_1^3W_{nn^{\prime}}}\theta\left(\frac{\tilde{\omega}_{fi}}{k_{nn^{\prime}}}\right)\eta_{\alpha},
\end{equation}
where $W_{nn^{\prime}}=(\tilde{\omega}_{fi}^2-k_{nn^{\prime}}^2)^{1/2}$ and $\eta_{\parallel}=W_{nn^{\prime}}^2/\tilde{\omega}_{fi}^2$ for the longitudinal dipole oscillations along the wave vector ${\bf k}_{nn¨{\prime}}$, $\eta_{\perp}=1$ for the transverse dipole oscillations in the lattice plane, and $\eta_z=k_{nn^{\prime}}^2/\tilde{\omega}_{fi}^2$ for the transverse dipole oscillations perpendicular to the lattice plane.\\
\section{Bose-Einstein condensation}\label{sec:BEC}
So far we have considered the lowest-energy excited states of plasmonic lattices which correspond to a single plasmon quantum in the system. The higher excited states can be described as an ensemble ("gas") of such quasiparticles which can "collide" with both each other and lattice phonons. The latter process can lead to their thermalization if the transition frequency between the adjacent excitonic levels is below the Debye frequency, $\omega_D$, of NPs which form the plasmonic lattice. For example, for silver NPs the Debye temperature $T_D=215$ K \cite{Smith95} that corresponds to $\omega_D=3.0\times 10^{13}$ s$^{-1}$ and this condition for $\epsilon_m=2.25$ is fulfilled if $(R/a)^3/N<10^{-3}$.\\
An additional requirement for establishing a thermal equilibrium is the dominance of the exciton-phonon interaction over the other relaxation channels, i.e. the thermalization rate should exceed both the radiative relaxation rate and the Landau damping rate \cite{Weick05}. \\
While the Landau damping can be suppressed in relatively large NPs \cite{Note1}, the radiative decay is inhibited for the excitonic states with $k>(\omega_{fi}/c)\sqrt{\epsilon_m}$. The latter states are not accessible for an incident light. Nevertheless they can be excited in the attenuated total reflection (ATR) configuration or with the use of a grating  which are common methods for the excitation of surface polaritons \cite{Raether88}. \\
Being initially excited in a non-radiative state, the plasmonic exciton gas can follow different scenarios of thermalization, depending on where the excitonic band energy minimum is located (see Fig. \ref{fig:energy}). If $V_{12}$ is negative, that occurs, for example, for the longitudinal dipole oscillations, the energy minimum lies at the center of the Brillouin zone ($k=0$) and the thermalization leads to the occupation of the rapidly decaying states in the radiative region. If $V_{12}$ is positive, that takes place, in particular, for the transverse dipole oscillations, the energy minimums are located  at the boundary of the first Brillouin zone ($k=\pm\pi/a$) and the thermalization leads to the occupation of the states for which the radiative relaxation is inhibited. In what follows, we consider the latter situation.\\
\subsection{1D lattice}\label{sec:BEC1}
The equilibrium populations of excitonic states follow the Bose-Einstein distribution \cite{Landau5}. We assume here that the number of NPs in the 1D lattice is large and the excitonic states form a quasi-continuous spectrum. The mean occupation number of the state with the energy $E$ is given by
\begin{equation}
\bar{n}(E)=\frac{1}{e^{(E-\mu)/k_BT}-1},
\end{equation}
where $\mu$ is the chemical potential, $k_B$ is the Boltzmann constant and $T$ is the lattice temperature. The total number of quasiparticles in the lattice of length $L$ is found as
\begin{equation}\label{eq:N}
\mathcal{N}=\frac{L}{2\pi}\int_{-\pi /a}^{\pi /a}\frac{dk}{e^{[E(k)-\mu ]/k_BT}-1},
\end{equation}
where $E(k)$ is given by Eq. (\ref{eq:Ek}). For $V_{12}>0$ the energy band (\ref{eq:Ek}) has two minimums $E_{min}=E_0-2V_{12}$ at $k=\pm \pi/a$.\\
Equation (\ref{eq:N}) implicitly determines the chemical potential of an ideal 1D "Bose gas" of plasmonic excitons in terms of its temperature and linear density $\mathcal{N}/L$. For a Bose gas the difference $E_{min}-\mu$ is always positive \cite{Landau5} and if the temperature of the gas is lowered at constant density, it decreases, tending to zero. The corresponding critical temperature $T_0$ is determined by the equation
\begin{equation}
\frac{\mathcal{N}}{L}=\frac{1}{\pi}\int_0^{k_N}\frac{dk}{e^{\epsilon(k)/k_BT_0}-1},
\end{equation}
where $\epsilon(k)=E(k)-E_{min}=2V_{12}(1+\cos ka)$ is the exciton energy counted from $E_{min}$. We have cut here the integral off at the wave vector of the lowest-energy state $k_N=N\pi/(N+1)a$ to avoid its divergence at $k=\pi/a$, that is a common problem for the density of states in low-dimensional systems \cite{Pethick08}. Taking into account that the main contribution to the integral comes from the region $k\approx k_N$, one obtains
\begin{equation}
T_0\approx \pi^2\frac{\mathcal{N}V_{12}}{N^2k_B},
\end{equation}
where we have used the relation $L/a\approx N$ and the inequality $N\gg 1$. For example, in the case of a chain of $N=20$ silver NPs with $R/d=0.2$ excited to the level of $\mathcal{N}=75$ excitons one finds $T_0\approx 320$ K. \\
When the lattice temperature is further decreased, the total number of quasiparticles in the "gas phase" is given by Eq. (\ref{eq:N}). The remaining $\mathcal{N}_c=\mathcal{N}-\mathcal{N}_g$ quasiparticles occupy the state with the lowest energy $E_N=\hbar\omega_1+2V_{12}\cos k_Na$ and form "a condensate" in the phase space, which is localized near the boundary of the first Brillouin zone. \\
As far as the condensate is located in the non-radiative region, it cannot radiate light. However the radiation is possible through the Umklapp process if a 1D lattice is disposed above a grating with the rulings oriented perpendicularly to it and the grating period $d_g$ satisfies the condition $d_g\approx 2a$, where we imply that the lattice constant $a$ is much less than the transition wavelength.\\
\subsection{2D lattice}
We consider in this section a 2D array of NPs occupying a rectangular area $S$. The total number of quasiparticles is given by
\begin{equation}\label{eq:N2}
\mathcal{N}=\frac{S}{(2\pi)^2}\int_{-\pi /a}^{\pi /a}\int_{-\pi /b}^{\pi /b}\frac{dk_xdk_y}{e^{[E({\bf k})-\mu ]/k_BT}-1}.
\end{equation}
where
\begin{equation}
E({\bf k})=E_0+2V_{12}(\cos k_xa + \cos k_yb).
\end{equation}
The critical temperature is found from the equation
\begin{equation}
\frac{\mathcal{N}}{S}=\frac{1}{\pi^2}\int_0^{k_{N_1}}\int_0^{k_{N_2}}\frac{dk_xdk_y}{e^{\epsilon({\bf k})/k_BT_0}-1}
\end{equation}
with $\epsilon({\bf k}) = 2V_{12}(2+\cos k_xa + \cos k_yb)$, $k_{N_1}=N_1\pi/(N_1+1)a$ and $k_{N_2}=N_2\pi/(N_2+1)b$.\\
A consideration similar to that given in Sec. \ref{sec:BEC1} leads to the following result:
\begin{equation}
T_0\approx  \frac{\pi}{2}\frac{\mathcal{N}V_{12}}{N_1N_2k_B}\left(\ln \frac{N_1N_2}{\sqrt{N_1^2+N_2^2}}\right)^{-1}.
\end{equation}
In this case the condensation occurs in the state characterized by the wave vector ${\bf k}_{N_1N_2}=(k_{N_1},k_{N_2})$ with the energy $E_{N_1N_2}=E_0+2V_{12}(\cos k_{N_1}a + \cos k_{N_2}b)$. Let us note that in a square array ($N_1=N_2$) with the same total number of NPs $N=N_1N_2$ and excitons $\mathcal{N}$ as in a 1D chain the critical temperature is $N/[\pi\ln (N/2)]$ times higher. \\
\section{Conclusion}\label{sec:conclusion}
In this paper, we have developed the theory of quantum plasmonic excitations in plasmonic lattices formed by spherical metal NPs. As a preliminary step, we have given a detailed derivation of the surface plasmon modes quantization in a single NP and obtained some basic results, which are necessary for the further discussion.\\ 
We have shown that the quantum excitations in a lattice are represented by the waves of the probability amplitude and found their energy spectrum in both 1D and 2D lattices. We noticed that such excitations when being initially localized at a certain NP (Frenkel-like plasmonic excitons) propagate across the lattice with the group velocity whose direction is controlled by the polarization of the exciting field. Having at hand the wave functions of the excited plasmonic states, we have calculated their radiative relaxation rates taking into account the action of a reflective substrate.\\
Tuning to the higher excited plasmonic states of a lattice, we introduced the concept of a Bose gas of plasmonic excitons. We have analyzed the conditions under which such a gas can be thermalized and found that it can exhibit the Bose-Einstein condensation at room temperature.\\
\appendix
\section{Explicit form of the tensors $\bar{G}^0$ and $\bar{G}^R$}\label{sec:apA}
The Fourier transforms $\bar{G}^{0,R}(z,z^{\prime};{\bm\kappa},\omega)$ have the following forms \cite{Sipe81,Sipe84}:
\begin{eqnarray}
\bar{G}^0(z,z^{\prime};{\bm\kappa},\omega)=2\pi i\tilde{\omega}^2W_m^{-1}\nonumber\\
\times\left[(\hat{\bf s}\hat{\bf s}+\hat{\bf p}_{0+}\hat{\bf p}_{0+})\theta(z-z^{\prime})e^{iW_m(z-z^{\prime})}\right.\nonumber\\
+\left.(\hat{\bf s}\hat{\bf s}+\hat{\bf p}_{0-}\hat{\bf p}_{0-})\theta(z^{\prime}-z)e^{-iW_m(z-z^{\prime})}\right]-4\pi\hat{\bf z}\hat{\bf z}\delta(z-z^{\prime})\nonumber\\
\end{eqnarray}
and
\begin{eqnarray}
\bar{G}^R(z,z^{\prime};{\bm\kappa},\omega)=2\pi i \tilde{\omega}^2W_m^{-1}\left(\hat{\bf s}\hat{\bf s}R_s+\hat{\bf p}_{0+}\hat{\bf p}_{0-}R_p\right)\nonumber\\
\times e^{i2W_mz_0}e^{iW_m(z-z^{\prime})},\nonumber\\
\end{eqnarray}
where $\tilde{\omega}=(\omega/c)\sqrt{\epsilon_m}$, $z_0$ is the distance between the NP chain and the substrate, $\hat{\bm \kappa}$, $\hat{\bf z}$ and $\hat{\bf s}=\hat{\bm \kappa}\times \hat{\bf z}$ are the unit vectors along the corresponding directions, $\hat{\bf p}_{0\pm}=\tilde{\omega}^{-1}(\kappa\hat{\bf z}\mp W_m\hat{\bm \kappa})$, $W_m=(\tilde{\omega}^2-\kappa^2)^{1/2}$, $\theta(x)$ is the unit step function, and $R_s$ and $R_p$ are the Fresnel reflection coefficients for $s$- and $p$-polarized light, respectively.\\
\section{Calculation of $R_{fi}$}\label{sec:apB}
After the substitution of the transition dipole moment into Eq. (\ref{eq:Rfi}) the transition rate takes the form
\begin{eqnarray}
R_{fi}=\frac{2\epsilon_m+1}{3}\omega_1R^3\frac{1}{N}\sum_{j,k=1}^Ne^{ik_n(x_j-x_k)}\nonumber\\
\times\frac{1}{2i(2\pi)^2}\sum_{\alpha,\beta}e_{\alpha}e_{\beta}\int\left[G_{\alpha\beta}(z_0,z_0;{\bf \kappa},\omega_{fi})e^{i\kappa_x(x_j-x_k)}\right.\nonumber\\
\left.-G^*_{\alpha\beta}(z_0,z_0;{\bf \kappa},\omega_{fi})e^{-i\kappa_x(x_j-x_k)}\right]d\kappa_xd\kappa_y,\nonumber\\
\end{eqnarray}
where it is implied that the $x$ axis is directed parallel to the chain of NPs so that all $y_j=0$ and all $z_j=z_0$. Assuming that the number of NPs is large, extending the summation over $j$ from $-\infty$ to $\infty$ and applying the Poisson summation formula
\begin{equation}
\sum_{j=-\infty}^{\infty}e^{i(k_n\pm \kappa_x)ja}=\frac{2\pi}{a}\sum_{p=-\infty}^{\infty}\delta\left(k_n\pm\kappa_x-\frac{2\pi}{a}p\right),
\end{equation}
one obtains
\begin{eqnarray}
R_{fi}\approx\frac{(2\epsilon_m+1) \omega_1R^3}{3a}\theta\left(\frac{\omega_{fi}\sqrt{\epsilon_m}}{ck_n}\right)\nonumber\\
\times\sum_{\alpha,\beta}e_{\alpha}e_{\beta}\int \text{Im}G_{\alpha\beta}(z_0,z_0;{\bm \kappa}_n,\omega_{fi})d\kappa_y
\end{eqnarray}
with ${\bm \kappa}_n=k_n\hat{\bf e}_x+\kappa_y\hat{\bf e}_y$, where we have taken into account that the terms with $p\neq 0$ give no contribution to $\text{Im}G_{\alpha\beta}$ if $a\ll 2\pi c/(\omega_{fi}\sqrt{\epsilon_m})$. Assuming here that the dipole moments of NPs are oriented along one of the coordinate axes, one finds
\begin{eqnarray}
R_{fi}^{\alpha}\approx\frac{(2\epsilon_m+1) \omega_1R^3}{3a}\theta\left(\frac{\omega_{fi}\sqrt{\epsilon_m}}{ck_n}\right)\nonumber\\
\times\int \text{Im}G_{\alpha\alpha}(z_0,z_0;{\bm \kappa}_n,\omega_{fi})d\kappa_y,
\end{eqnarray}
where $\alpha=x,y,z$.\\

\end{document}